\documentclass[prd,aps,notitlepage,twocolumn,longbibliography]{revtex4-2} 
\usepackage{amsfonts,amssymb,amsmath,mathrsfs,graphicx,xcolor,bm,mathtools,orcidlink,url}
\definecolor{ultramarine}{rgb}{0.07, 0.04, 0.56}
\definecolor{cadmiumgreen}{rgb}{0.0, 0.42, 0.24}
\definecolor{indigo(dye)}{rgb}{0.0, 0.25, 0.42}
\usepackage{hyperref}
\hypersetup{
colorlinks=true,
citecolor=ultramarine,
linkcolor=cadmiumgreen,
urlcolor=indigo(dye),
}
\newcommand{\f}[2]{\frac{#1}{#2}}
 
\newcommand{\mk}[1]{\left( #1 \right)}  
\newcommand{\kk}[1]{\left[ #1 \right]}  

\newcommand{\sk}[1]{(\tilde #1 )}
\newcommand{\be}{\begin{equation}}  
\newcommand{\ee}{\end{equation}}
\newcommand{\dd}{\mathop{}\!\mathrm{d}}

\newcommand{\Cref}{C^{\rm ref}}
\newcommand{\mA}{\mathcal{A}}
\newcommand{\mC}{\mathcal{C}}
\newcommand{\mE}{\mathcal{E}}
\newcommand{\mO}{\mathcal{O}}
\newcommand{\mW}{\Delta}
\newcommand{\pa}{\partial}

\newcommand{\Ain}{A_{\rm in}}
\newcommand{\Aout}{A_{\rm out}}

\newcommand{\secl}[1]{\textbf{#1—}}

\begin{document}

\title{
Resonant Excitation of Quasinormal Modes of Black Holes 
}

\author{Hayato Motohashi\,\orcidlink{0000-0002-4330-7024}}
\thanks{Present address: Department of Physics, Tokyo Metropolitan University, 1-1 Minami-Osawa, Hachioji, Tokyo 192-0397, Japan}
\affiliation{Division of Liberal Arts, Kogakuin University, 2665-1 Nakano-machi, Hachioji, Tokyo, 192-0015, Japan}


\begin{abstract}
We elucidate that a distinctive resonant excitation between quasinormal modes (QNMs) of black holes emerges as a universal phenomenon at an avoided crossing near the exceptional point through high-precision numerical analysis and theory of QNMs based on the framework of non-Hermitian physics. 
This resonant phenomenon not only allows us to decipher a long-standing mystery concerning the peculiar behaviors of QNMs but also stands as a novel beacon for characterizing black hole spacetime geometry.  
Our findings pave the way for rigorous examinations of black holes and the exploration of new physics in gravity.
\end{abstract}

\maketitle  


\secl{Introduction}
The last decade has witnessed the advent of gravitational wave (GW) astronomy, opening a new window to the universe, especially in the dynamical and strong-field regime near black holes~\cite{Abbott:2016blz,LIGOScientific:2014pky,VIRGO:2014yos,KAGRA:2020tym}. 
General relativity predicts that black holes are the simplest astrophysical objects in the universe since the only rotating astrophysical black hole solution that satisfies the Einstein equation is the Kerr spacetime~\cite{Israel:1967wq,Carter:1971zc,Hawking:1971vc}, characterized solely by mass and spin.
Remnant black holes after binary black hole mergers are expected to settle into the state of a Kerr black hole, emitting ringdown GWs. 
Linear perturbation theory predicts that the ringdown GWs are described by a linear combination of intrinsic exponentially damped sinusoids known as quasinormal modes (QNMs)~\cite{Vishveshwara:1970zz,Press:1971wr,Teukolsky:1973ha,Chandrasekhar:1975zza}.
Their timescales and amplitudes are governed by complex frequencies and excitation factors~\cite{Leaver:1985ax,Leaver:1986a,Leaver:1986gd,Sun:1988tz,Nollert:1992ifk,Nollert:1998ys,Andersson:1995zk,Andersson:1996cm,Glampedakis:2001js,Glampedakis:2003dn}, also determined solely by mass and spin, playing a central role in the black hole spectroscopy program~\cite{Detweiler:1980gk,Echeverria:1989hg,Finn:1992wt,Dreyer:2003bv,Berti:2005ys}. 
Despite extensive efforts to better understand them, the peculiar behaviors of QNMs under changes in black hole parameters have remained elusive~\cite{Kokkotas:1999bd,Nollert:1999ji,Berti:2009kk,Konoplya:2011qq}.

In this Letter, we reveal the emergence of the resonant excitation phenomenon associated with an avoided crossing of QNMs near the exceptional point (EP).
The avoided crossing in quantum mechanics, also known as level repulsion, is a phenomenon where two energy eigenvalues cannot coincide unless a certain condition is satisfied~\cite{Hund1927,1929PhyZ...30..467V,LandauQM,Arnold1978}.
It plays a pivotal role in experiments and observations across vast fields of physical science~\cite{ashcroft1976solid,1932PhyZS...2...46L,1932RSPSA.137..696Z,Majorana1932,Stu1932,Ivakhnenko:2022sfl,Purcell1946,Herzberg1991,Smirnov:2003da,Wurm:2017cmm,Giganti:2017fhf}.
The EP~\cite{Kato1995,2019NatMa..18..783O,Wiersig:20,Parto2021,Ding:2022juv}, a branch point singularity in the complex eigenvalue plane, is crucial in non-Hermitian physics~\cite{El-Ganainy2018,Bergholtz:2019deh,Ashida:2020dkc}, where avoided crossing occurs in its vicinity~\cite{PhysRevE.61.929,Heiss:2012dx}.
With these concepts, we formulate the resonance of QNMs and provide novel perspectives for exploring gravity.
We employ geometric units with $G=c=1$ throughout.

\secl{Kerr QNM frequencies and excitation factors}
We first reveal the emergence of the resonance of Kerr QNMs at avoided crossing and identify their unique features. 
Linear perturbation theory in general relativity predicts that the ringdown GW strain $h=h_+-ih_\times$ at a far distance from a perturbed Kerr black hole is given by a superposition of QNM damped sinusoids~\cite{Vishveshwara:1970zz,Press:1971wr,Teukolsky:1973ha,Chandrasekhar:1975zza}:
\begin{align} h = \sum_{\ell, m, n} \f{I_{\ell mn} B_{\ell mn} S_{\ell mn}e^{im\varphi}}{\omega_{\ell mn}^{2}}e^{-i\omega_{\ell mn}t} , \end{align}
in a decomposition in terms of spin-weighted spheroidal harmonics $S_{\ell mn}(\theta,a\omega_{\ell mn})$~\cite{Teukolsky:1972my,Teukolsky:1973ha,Press:1973zz}.
While a complex factor $I_{\ell mn}$ involves an integral depending on the initial condition, recent studies suggest that for each $(\ell,m)$ multipole $I_{\ell mn}$ does not strongly depend on the overtone index $n$ or black hole spin~\cite{Oshita:2021iyn,Cheung:2023vki}.
Therefore, the ringdown GWs are mainly governed by complex QNM frequencies 
$\omega_{\ell mn}$
and excitation factors $B_{\ell mn}$ defined by~\cite{Leaver:1986gd}
\begin{align} \label{Bdef} B_{\ell mn} = \left. \f{\Aout}{2\omega} \mk{\f{\dd\Ain}{\dd\omega}}^{-1} \right|_{\omega=\omega_{\ell mn}} , \end{align}
with asymptotic amplitudes $\Aout$ and $\Ain$ for the outgoing and ingoing waves, respectively.
Both $\omega_{\ell mn}$ and $B_{\ell mn}$ depend only on mass $M$ and spin $a$ of the black hole.
Since the mass dependency is a simple scaling, the dependency on a dimensionless spin parameter $a/M$ is of primary interest.
Except for $a/M$, we normalize physical quantities by $2M$.

\begin{figure*}[t]
  \centering
  \includegraphics[width=0.33\textwidth]{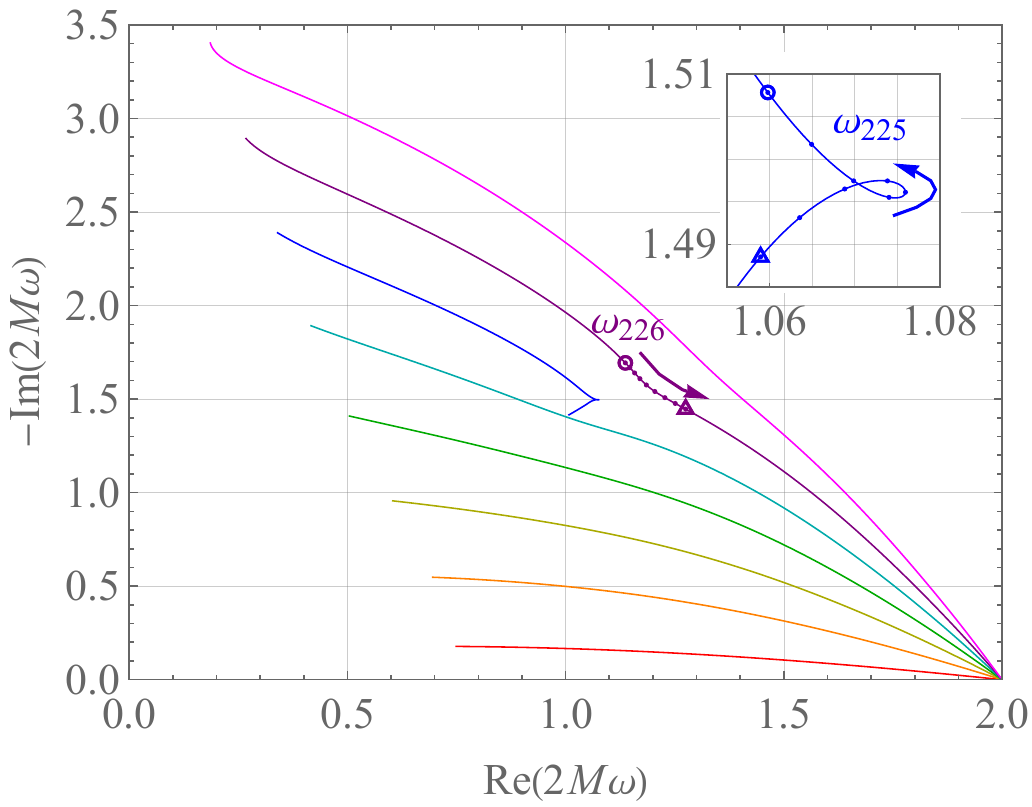}
  \includegraphics[width=0.42\textwidth]{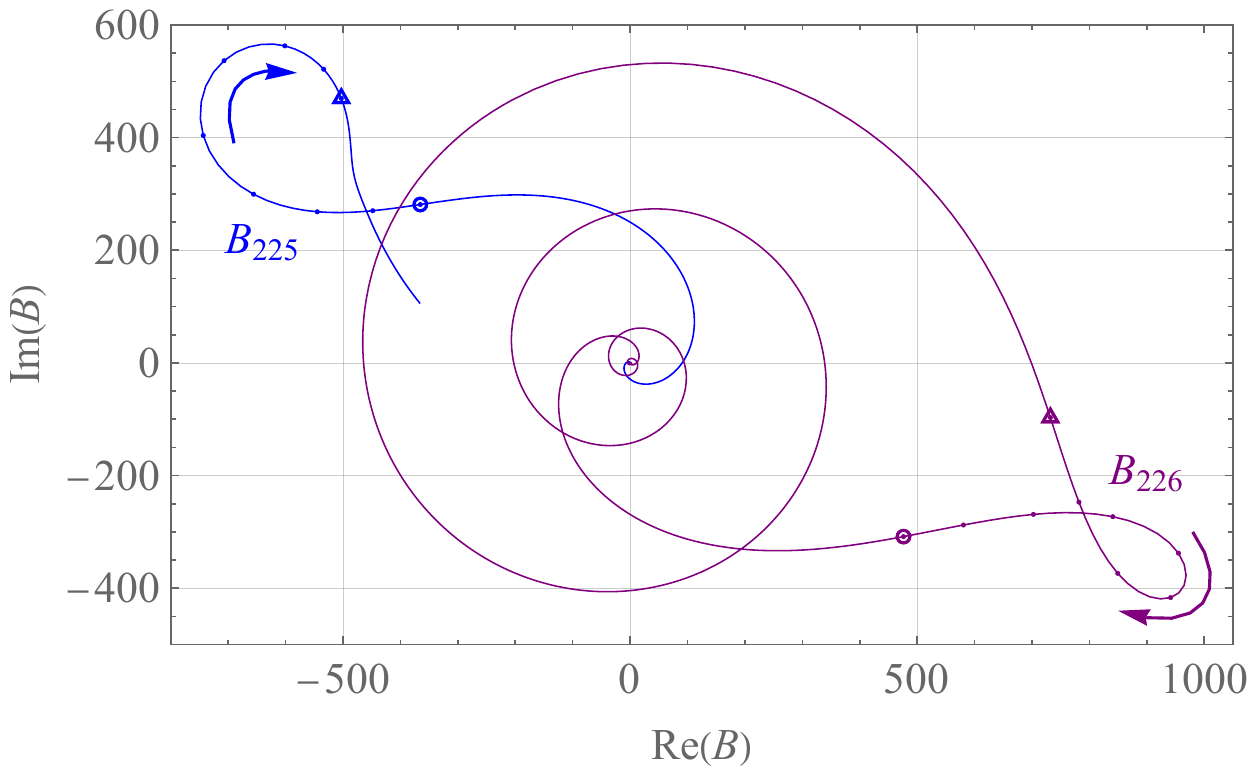}
  \includegraphics[width=0.235\textwidth]{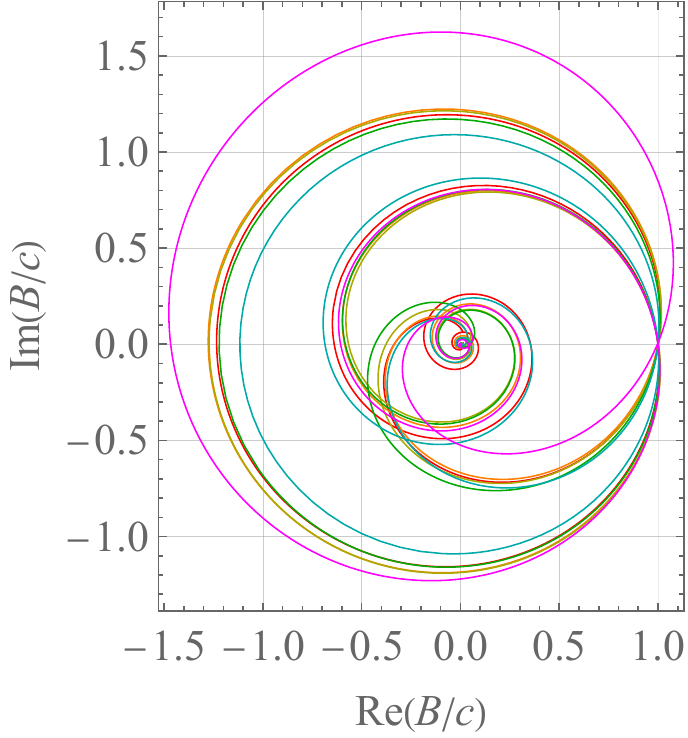}
\caption{Kerr QNM frequencies $\omega_{\ell mn}$ (left) of $(\ell,m)=(2,2)$ from $n=0$ (bottom red) to $7$ (top magenta) for $0\leq a/M\leq 1-10^{-6}$ with arrows indicating the direction of increasing spin.
Excitation factors $B_{\ell mn}$ with $n=5,6$ (middle) and other overtones (right), the latter of which are normalized by a numerically chosen complex constant $c_n$ to highlight the similarity of the spiral shapes.
The range $a/M=0.875$ (circle)—$0.915$ (triangle) is highlighted with small dots with a spacing of $0.005$.}
\label{fig:wB22n-zoom}
\end{figure*}

\begin{figure}[t]
  \includegraphics[width=0.78\columnwidth]{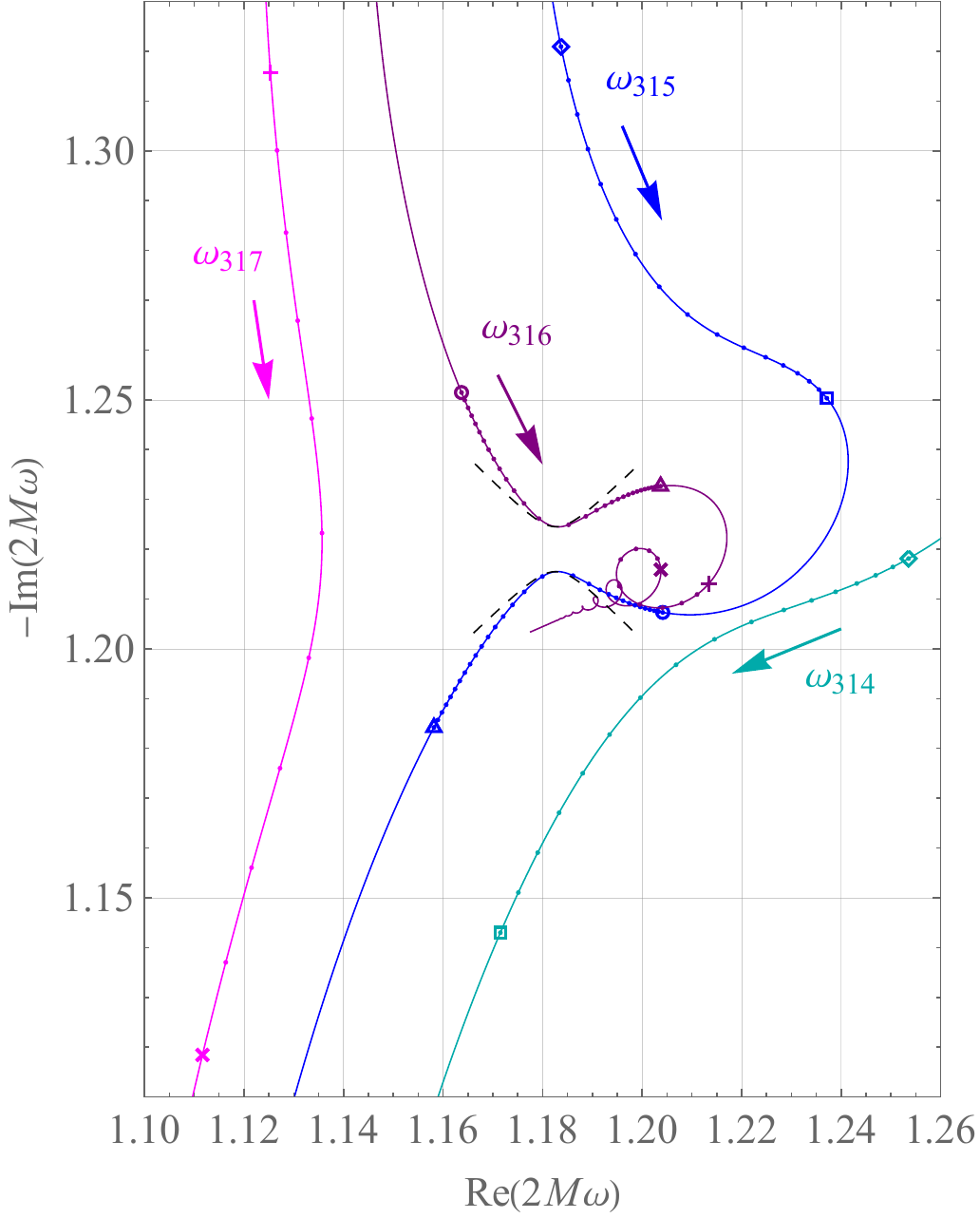}
  \includegraphics[width=0.83\columnwidth]{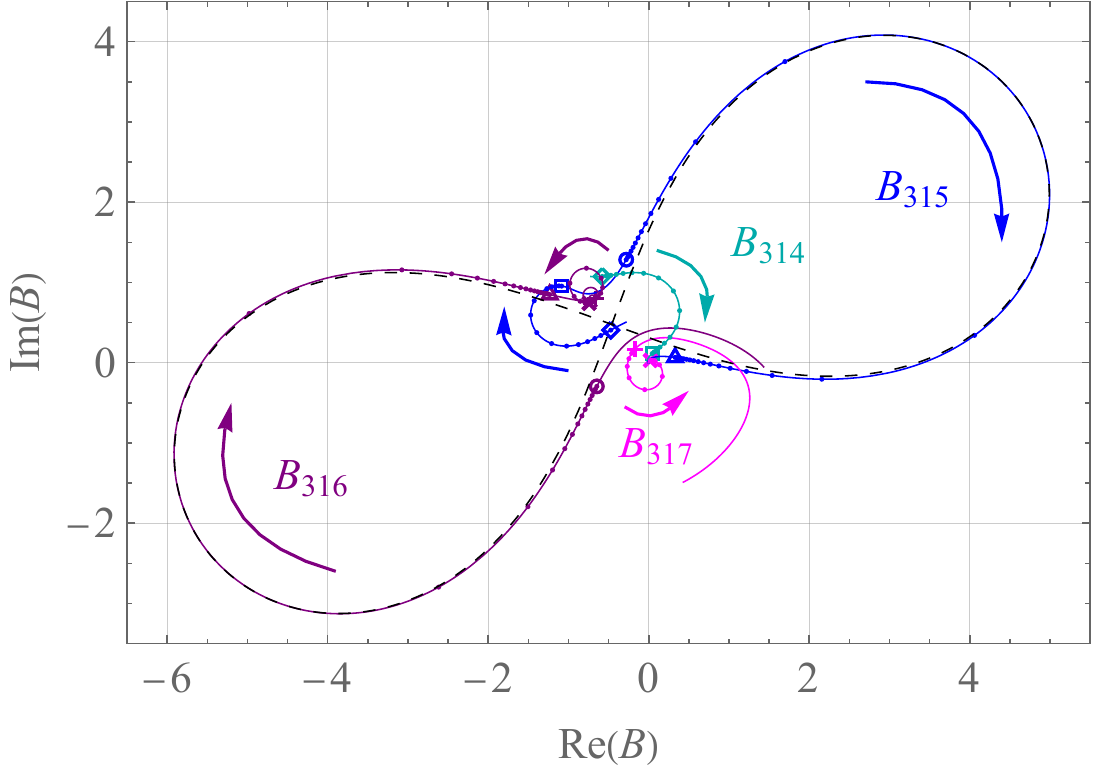}
  \caption{Kerr QNM frequencies $\omega_{\ell mn}$ (top) and excitation factors $B_{\ell mn}$ (bottom) of $(\ell,m)=(3,1)$ from $n=4$ to $7$ for $0.95 \leq a/M \leq 1-10^{-6}$. 
  Highlighted are the ranges $a/M=0.952$ (diamond)—$0.96$ (square), $0.9715$ (circle)—$0.973$ (triangle), and $0.978$ (plus)—$0.983$ (cross) with small dots with a spacing of $5\times 10^{-4}$, $5\times 10^{-5}$, and $5\times 10^{-4}$, respectively.
  Black dashed curves represent the hyperbola (top) and lemniscate (bottom).}
  \label{fig:wB31n}
\end{figure}

We revisit numerical calculation of the QNM frequencies and excitation factors of Kerr black holes.
We closely follow the strategy adopted in Ref.~\cite{Zhang:2013ksa} and implement some technical improvements to achieve high-precision calculation up to higher overtones
which will be presented elsewhere~\footnote{In addition, we correct errors found in an intermediate step, Eqs.~(27)--(29), of Ref.~\cite{Zhang:2013ksa}, which appear to have originated from errors in the first version of Ref.~\cite{Sasaki:2003xr}. Consequently, our results of excitation factors and those in Ref.~\cite{Zhang:2013ksa} differ by a factor of $e^{-i\omega (1 - \sqrt{1 - (a/M)^2})}$.}.
We obtain results for gravitational, electromagnetic, and scalar perturbations~\cite{motohashi_2024_12696858}.

For the $(\ell,m)=(2,2)$ multipole of GWs, most Kerr QNM frequencies uniformly migrate towards the accumulation point at $2M\omega_{\ell mn}=m$~\cite{Detweiler:1980gk,Glampedakis:2001js,Cardoso:2004hh,Hod:2008zz,Yang:2012pj,Yang:2013uba}.
However, as depicted in Fig.~\ref{fig:wB22n-zoom}, only the fifth overtone suddenly turns over at $a/M\simeq 0.9$, forming a small knot-shaped loop and moving towards an isolated point far from the accumulation point.
This anomaly, akin to an unexpected dissonance in music, was first identified by Onozawa~\cite{Onozawa:1996ux} about three decades ago and later confirmed by other work~\cite{Berti:2004md,Cook:2014cta}, but the physical reason has been veiled.
While it has been implicitly assumed that the fifth overtone solely behaves anomalously,
here we point out that, while less manifest, the trajectory of the sixth overtone QNM frequency is slightly distorted when it passes near the fifth overtone at $a/M\simeq 0.9$.

On the other hand, it was recently found that the absolute values of $B_{225}$ and $B_{226}$ become large for high-spin Kerr black holes~\cite{Giesler:2019uxc,Oshita:2021iyn}, despite the fact that the excitation factors tend to zero at the extremal spin~\cite{Ferrari:1984zz,Berti:2006wq,Zhang:2013ksa} and asymptotically scale as $n^{-1}$ for higher overtones~\cite{Andersson:1996cm,Berti:2006wq}. 
Again, the physical reason has remained unclear.
However, when viewed on the complex plane in Fig.~\ref{fig:wB22n-zoom}, we find that the excitation factors actually exhibit a suggestive behavior:
While other modes follow spiral trajectories, it turns out that $B_{225}$ and $B_{226}$ are against the spirals and are enhanced in almost opposite directions at $a/M\simeq 0.9$.

The implications of these observations are twofold: 
First, this phenomenon should be understood as occurring in a pair of modes rather than a single mode. 
Second, there should be an underlying connection between the turning or distortion of the QNM frequencies and the strong excitations.

Actually, a similar phenomenon occurs more manifestly for higher multipoles.
In Fig.~\ref{fig:wB31n}, we see that $(\ell,m)=(3,1)$ modes exhibit the distinctive behaviors successively between multiple pairs.
Among them, the pair of $n=4$ and $5$ is the most manifest case, exhibiting a sharp repulsion between the two QNM frequencies~\footnote{Such repulsion was also observed in the QNMs of scalar field in Kerr~\cite{Yang:2013uba}, (anti-)de Sitter black holes~\cite{Jansen:2017oag,Davey:2022vyx,Kinoshita:2023iad}, and charged black holes~\cite{Dias:2021yju,Dias:2022oqm,Davey:2024xvd}.}.
In Fig.~\ref{fig:wB31n}, we also note that the mild repulsion between $n=6$ and $7$ overtones seems to be a trigger of the looping trajectory of the sixth overtone observed in Refs.~\cite{Andersson:1996xw,Onozawa:1996ux}.
Further, during these peculiar behaviors of QNM frequencies, the corresponding pairs of excitation factors exhibit symmetric amplifications.
The sharper the repulsion between two QNM frequencies is, the more strongly and more clearly in a point-symmetric manner the corresponding pair of excitation factors is amplified.

Interestingly, for the most manifest case in Fig.~\ref{fig:wB31n},
we find that the QNM frequencies and excitation factors almost follow the hyperbola and the lemniscate of Bernoulli, respectively.
When written down in the polar coordinates with $z=re^{i\theta}$, the hyperbola $r^{-2}=\cos 2\theta$ and the lemniscate $r^2=\cos 2\theta$ are inverses of each other.
Albeit approximate, it is remarkable that such intimately related curves are inherent in the QNM spectrum of Kerr GWs.

\begin{figure*}[t]
\centering
\includegraphics[width=.245\textwidth]{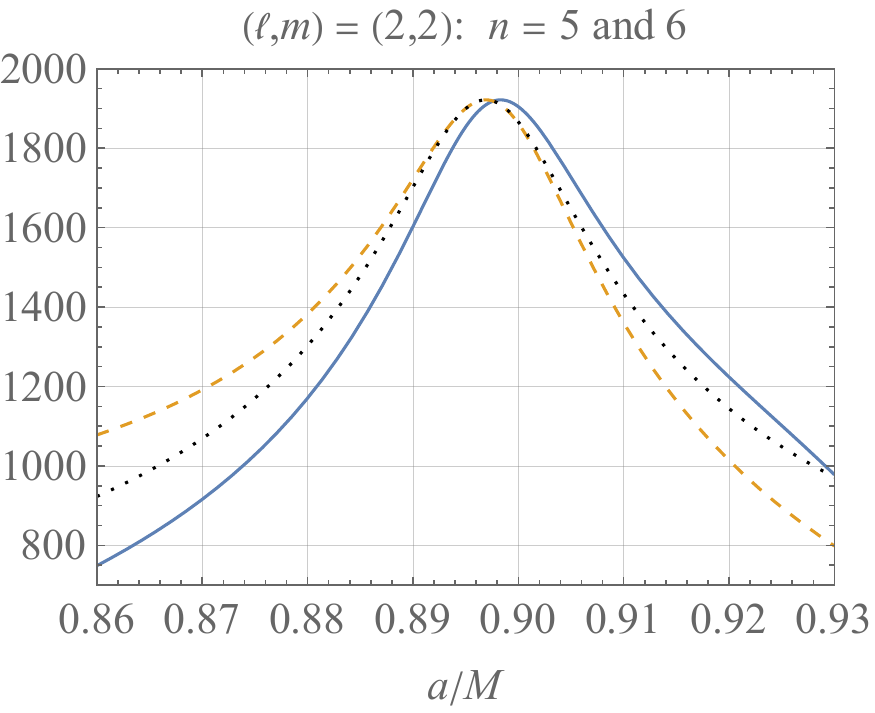}
\includegraphics[width=.245\textwidth]{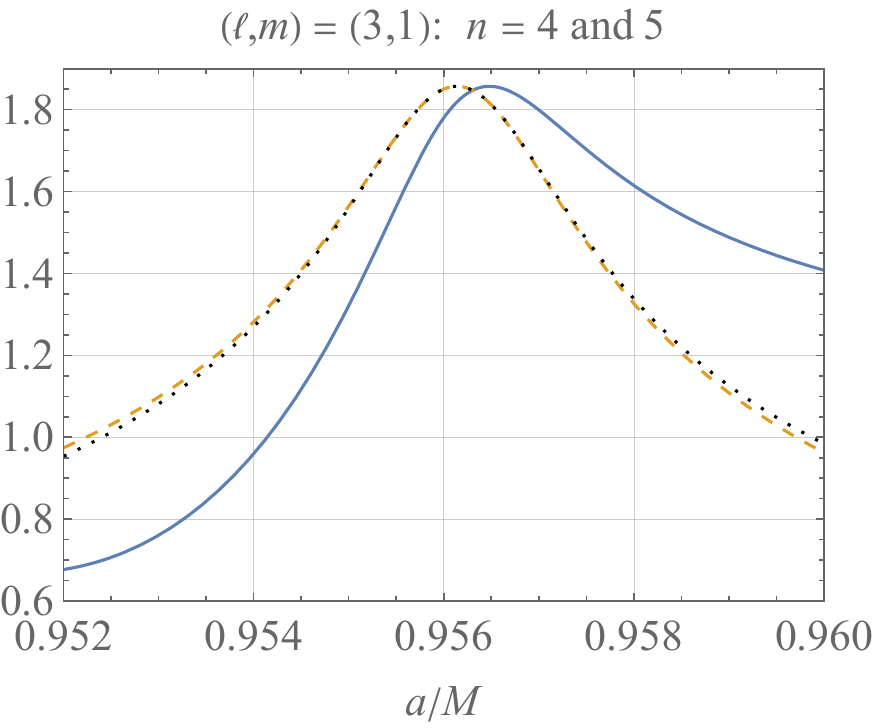}
\includegraphics[width=.245\textwidth]{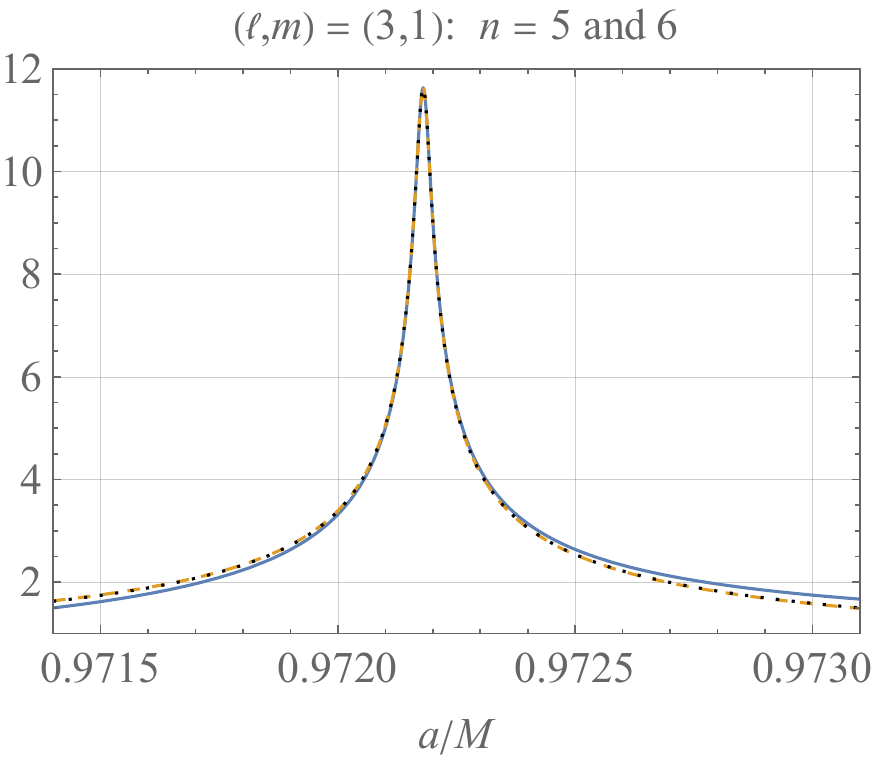}
\includegraphics[width=.245\textwidth]{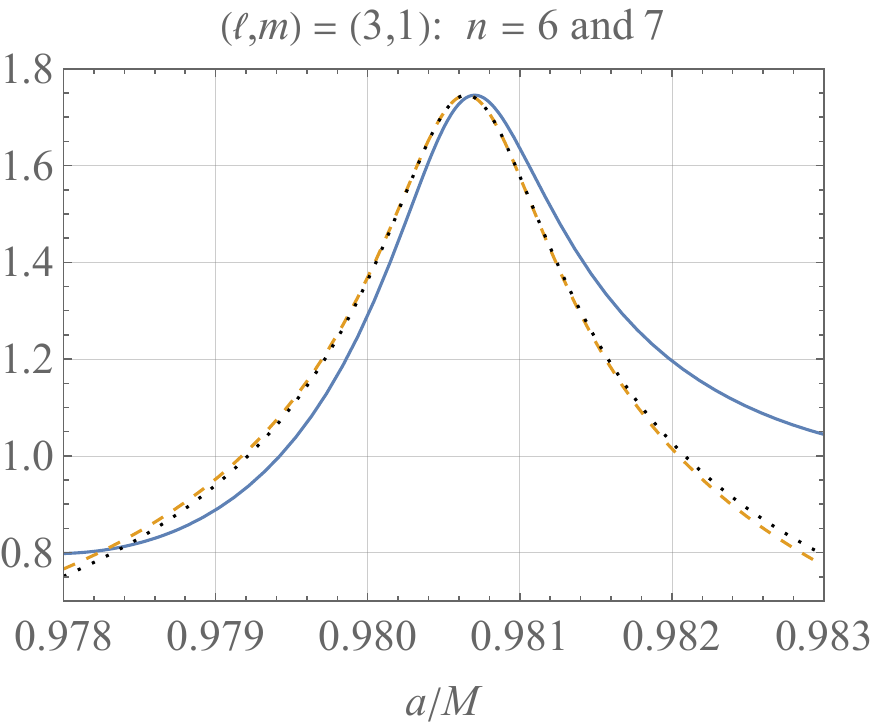}
\caption{Absolute values of the difference between the excitation factors (blue solid) and the inverse of the difference between QNM frequencies (orange dashed), the latter of which is normalized to match the maxima, with a fit by a quarter-power Lorentzian (black dotted).}
\label{fig:DBDw}
\end{figure*}

We show in Fig.~\ref{fig:DBDw} that the inverse proportionality between the difference of excitation factors $\Delta B$ and the difference of QNM frequencies $\Delta \omega$ is actually a common feature of the pair excitations, providing insight that they belong to the same underlying phenomenon.
As expected, the inverse proportionality is more manifest for the case of sharper repulsion and stronger excitation.
Further, we find that the peaks in Fig.~\ref{fig:DBDw}, especially the sharp ones, demonstrate an excellent agreement with a quarter-power Lorentzian $f(a/M)$, 
where $f(x) = f_0 [ \gamma^2/\{ (x-x_0)^2 + \gamma^2 \} ]^{1/4}$ with positive parameters $f_0, \gamma, x_0$, 
rather than the Lorentzian itself.
Furthermore, we also confirm that a product $\Delta B \Delta \omega$ remains almost constant as a complex number during the resonant excitation.

We find that this phenomenon ubiquitously occurs for Kerr QNMs, i.e., GWs with other multipoles, and scalar and electromagnetic fields~\cite{motohashi_2024_12696858}.
Further, even in QNMs apart from Kerr black holes, we find that the resonances with the same features show up~\footnote{See \hyperref[sec:supp]{Supplemental Material}, which includes Refs.~\cite{Barausse:2014tra,Cheung:2021bol,Kyutoku:2022gbr,Jona-Lasinio1981,Graffi_1984,SIMON1985123,Santarsiero_2019,Nollert:1996rf,Daghigh:2020jyk,Qian:2020cnz,Jaramillo:2021tmt,Berti:2022xfj,Cardoso:2024mrw}, for an additional analysis of the resonance phenomenon between QNMs of a toy model.}.
These are tantalizing hints to a universal nature of QNMs.

\secl{Theory of quasinormal modes}
Below, we provide a theoretical explanation for this phenomenon by developing a theory of QNMs.
Let us consider the QNMs of a general 
master equation for perturbations $\psi$ around a black hole after Fourier and harmonic expansions,
\begin{align} \label{gen-Sch-eq} \mk{\f{\dd^2}{\dd x^2} + U }\psi=0, \end{align}
where $U$ is an effective potential and $x$ denotes the tortoise coordinate with $x\to\pm\infty$ corresponding to the spatial infinity and event horizon of the black hole, respectively.
For nonrotating black holes, $U = \omega^2-V(x)$~\cite{Regge:1957td,Zerilli:1970se} and hence Eq.~\eqref{gen-Sch-eq} reads $H\psi = \omega^2\psi$ with an effective Hamiltonian $H = -\f{\dd^2}{\dd x^2} + V(x)$, which is analogous to the Schr\"odinger equation in quantum mechanics.
For the radial perturbation equation of rotating black holes, $U$ depends on $x, \omega$ and other parameters~\cite{Teukolsky:1972my,Teukolsky:1973ha}, so Eq.~\eqref{gen-Sch-eq} defines a nonlinear eigenvalue problem.
$U$ changes as the spin parameter varies, causing the migration of QNM frequencies.

A crucial difference of QNMs from the bound state in quantum mechanics originates from the boundary condition.
Instead of decaying at infinity, the QNM wave function is required to satisfy the ingoing or outgoing wave condition $\psi\to e^{\pm i\omega x}$ as $x\to \pm \infty$.
This Siegert boundary condition~\cite{Siegert:1939zz} enforces QNM eigenvalues to be discrete complex numbers with negative imaginary parts.
As a result, the QNM wave function diverges at $x\to \pm \infty$ (when viewed on a constant time hypersurface), and 
the standard inner product for bound states is not well-defined for QNMs.
Several kinds of definitions of the inner product for QNMs of black holes have been considered~\cite{Ching:1993gt,Leung:1997was,Leung_1998,Yang:2014tla,Zimmerman:2014aha,Jaramillo:2020tuu,Gasperin:2021kfv,Yang:2022wlm,Green:2022htq,London:2023aeo,London:2023idh,Ma:2024qcv}. 

In quantum mechanics, this kind of system is known as the Schr\"odinger problem of open systems and has been utilized for studies of quantum resonances, for which the Hamiltonian is not Hermitian~\cite{Gamow:1928zz,LandauQM,Kukulin1989,moiseyev_2011}.
More recently, the non-Hermitian physics has been extensively studied both from a theoretical and experimental point of view~\cite{El-Ganainy2018,Bergholtz:2019deh,Ashida:2020dkc}. 
Here, we introduce a regular inner product, or more precisely biorthogonal product, for QNMs, inspired by the case of the quantum resonance states~\cite{Kukulin1989,PhysRevA.4.1782,PhysRevA.3.1217,PhysRevC.6.114}. 
It corresponds to a generalization of Refs.~\cite{Ching:1993gt,Leung:1997was,Leung_1998} to the case with Eq.~\eqref{gen-Sch-eq} including rotating black holes. 

In this formulation, the QNM wave function $\psi_n$ with eigenvalue $\omega_n$ corresponds to the resonance state, for which we introduce the so-called antiresonance (or adjoint) state $\tilde\psi_n\coloneqq \psi_n^*$ with eigenvalue $\tilde \omega_n \coloneqq -\omega_n^*$.
We then define the biorthogonal product for QNMs along a contour $\mC$ as
\begin{align} \sk{\psi_n | \psi_k} 
\coloneqq \int_\mC \dd{x} \f{U_n-U_k}{\omega_n^2-\omega_k^2} \psi_n\psi_k - \f{[ \psi_n\psi'_k - \psi'_n\psi_k]_{\pa\mC}}{\omega_n^2-\omega_k^2} ,
\end{align}
where $U_n=U|_{\omega=\omega_n}$.
From Eq.~\eqref{gen-Sch-eq}, it is straightforward to show that the biorthogonality exactly holds for any $\mC$, i.e., 
$\sk{\psi_n|\psi_k} = 0$ when $n\ne k$.
Formally taking the limit $\omega_k\to \omega_n$, we define the ``norm squared'' for QNMs as
\begin{align} 
\label{reg-2n} &\mA_{n}^{2} \coloneqq 
\int_\mC \dd{x} \f{\pa U_n}{\pa\omega_n^2}\psi_n^2 - 
\left.\f{\pa}{\pa \omega_k^2}[ \psi'_n\psi_k - \psi_n\psi'_k ]_{\pa\mC} \right|_{\omega_k=\omega_n} ,
\end{align}
which is in general complex~\footnote{As a special case, if we choose a contour $\mC$ along the real axis, \eqref{reg-2n} reduces to the Zel'dovich regularization~\cite{Zeldovich:1961a} (see also earlier work~\cite{Kapur1938}).}. 
We can analytically show that $\mA_n^2$ satisfies the following useful formula~\cite{Motohashi:prep}:
\begin{align} \label{Bana} B_n = \f{i\Aout^2}{2\omega_n \mA_n^2} , \end{align}
which allows for an alternative way to calculate the excitation factor $B_n$ defined by Eq.~\eqref{Bdef}.
The agreement between Eqs.~\eqref{Bana} and \eqref{Bdef} is numerically crosschecked for various specific cases, including Schwarzschild and Kerr black holes with gravitational, electromagnetic, and scalar perturbations across various spin parameters.
The formula~\eqref{Bana} reveals that the excitation factor $B_n$ becomes large when the norm squared $\mA_n^2$ becomes small, which precisely occurs at the avoided crossing near the EP.

\secl{Resonant excitation at avoided crossing}
To investigate how QNM frequencies and excitation factors change when two QNM frequencies get close due to the variation of $U$, let us regard the continuous variation as an accumulation of infinitesimal changes $\delta U$.
If a mode $\omega_n$ is far from degeneracy, it is straightforward to generalize the Rayleigh-Schr\"odinger-like perturbation theory for resonance state~\cite{Zeldovich:1961a,PeZe1998} to QNMs to obtain~\cite{Motohashi:prep}
\begin{align} \label{dw-reg} 
\delta\omega_n = -\f{ \sk{\psi_n | \delta U_n | \psi_n} }{ 2\omega_n\sk{\psi_n | \psi_n} }
= iB_n \f{\sk{\psi_n | \delta U_n | \psi_n}}{\Aout^2},
\end{align}
which reveals another role of the excitation factor $B_n$: defining the sensitivity of $\delta\omega_n$ to $\delta U$.

The avoided crossing, where two (or more) eigenvalues are close to degenerate, requires special care~\footnote{The similarity between the avoided crossing and repulsion of QNM eigenvalues was pointed out in Refs.~\cite{Dias:2021yju,Dias:2022oqm,Davey:2024xvd} in studies of charged black holes, where a formula similar to, but distinct from, Eq.~\eqref{wpm} was derived, and it was claimed that such repulsion does not occur for Kerr QNMs.
Here, we examine QNM frequencies and also excitation factors, and derive the avoided crossing and resonant excitation as a universal phenomenon including the Kerr case.}.
We employ a simple analysis of the bifurcation theory of eigenvalues about the EP~\cite{PhysRevE.61.929,Seyranian2003} to capture the essential features of the resonance.
We focus on the nonrotating black hole case with $U=\omega^2-V(x)$~\footnote{The rotating black hole case may be investigated by applying a prescription of auxiliary eigenvalue recently employed in \cite{PhysRevLett.132.126601} in the context of the bulk-edge correspondence.} and assume that two QNM overtones are close, $\omega_{n_1}\approx \omega_{n_2}$, and are isolated from other QNMs so that we can approximately treat them as a two-level system.
Suppose a small change $V \to V+\delta V$ as a part of continuous variation, causing a shift of the QNM frequencies $\omega_{n_i} \to \omega_\pm$, which can be obtained as eigenvalues of the Hamiltonian $H_{n_in_j} = \sk{\psi_{n_i}|H|\psi_{n_j}}/(\mA_{n_i} \mA_{n_j})$:
\begin{align} \label{wpm} \omega^2_\pm = \mE_c \pm \sqrt{\mE_d^2+\mW^2} , \end{align}
where $\mE_{c,d} \coloneqq (\mE_{n_1}\pm \mE_{n_2})/2$,  
$\mE_{n_i} \coloneqq \omega_{n_i}^2 + \delta\omega_{n_i}^2$ with $i=1,2$, and $\mW \coloneqq \sk{\psi_{n_1}|\delta V|\psi_{n_2}} /(\mA_{n_1} \mA_{n_2})$.
When $\mW$ is negligible, Eq.~\eqref{wpm} recovers $\mE_{n_i}$, the result of perturbation theory. 
On the other hand, $\omega_\pm$ are degenerate if $\mE_d^2+\mW^2=0$ is satisfied, which defines an EP.
Since this condition involves complex variables, it is in general not possible to be satisfied by tuning a single real parameter, resulting in the avoided crossing.

Let us consider the case where continuous variation of $V$ is driven by a single parameter. 
At each point except the avoided crossing the perturbation theory is valid, and hence $\omega_{n_i}^2$ follow $\mE_{n_i}$.
Suppose that the perturbation theory predicts that the two modes approach each other along the same line, for which, denoting $\mE_d=pe^{i\alpha}$ and $\mE_c=\omega_{\rm EP}^2$, $p$ is a real parameter and $\alpha$ and $\omega_{\rm EP}^2$ are constant.
We also neglect the variation of $\mW=qe^{i\beta}$. 
The single-parameter degree of freedom of $V$ is thus represented by $p$.
Denoting $\omega_\pm^2-\omega_{\rm EP}^2 = x+iy$, it is straightforward to derive
$X^2-Y^2 = q^2\sin 2(\alpha-\beta)$, 
where $X+iY$ is defined by a rotation of $x+iy$ by $\pi/4-\alpha$. 
Unless $\alpha=\beta$, which would require tuning an additional parameter, $\omega_\pm^2-\omega_{\rm EP}^2$ follows a hyperbolic avoided crossing, and so does $\omega_\pm - \omega_{\rm EP}$ itself, since this occurs at a small region near the EP.
A simple calculation shows $|\omega_\pm^2 - \omega_{\rm EP}^2| \approx ({\rm const}+p^2)^{1/4}$ at the peak when $p\ll 1$, meaning that $|\omega_+ - \omega_- |^{-1}$ obeys the quarter-power Lorentzian.

On the other hand, QNM eigenstates $\psi_\pm$ associated with $\omega_\pm$ are given by 
$\psi_+ = \psi_{n_1} + T\psi_{n_2}$ and  
$\psi_- = -T\psi_{n_1} + \psi_{n_2}$,
where $T=1/(\delta+\sqrt{\delta^2-1})$ is governed by a dimensionless complex parameter 
$\delta = \sqrt{\mE_d^2/\mW^2+1}$.
If the two modes are close and then $\delta\approx 0$ and $\mA_{n_1}^2\approx \mA_{n_2}^2$ are satisfied, the eigenstates are almost degenerate $\psi_-\approx i\psi_+$, and the norm squared nearly vanishes as $\mA^2_\pm\propto \pm\delta$.
These behaviors are nothing but what is expected at the EP~\cite{2019NatMa..18..783O,Wiersig:20,Parto2021,Ding:2022juv}.
With Eqs.~\eqref{Bana} and \eqref{wpm}, we obtain $B_\pm\propto \mA^{-2}_\pm\propto (\omega_\pm^2-\omega_{\rm EP}^2)^{-1}$. 
Hence, the excitation factors are amplified following the inverse of the hyperbola, i.e., the lemniscate of Bernoulli, and the resonance peak obeys the quarter-power Lorentzian.

\secl{Conclusions and Discussion}
When two QNM frequencies approach each other on the complex plane, they exhibit anomalously amplified excitations. 
We have elucidated this resonance phenomenon near EPs by developing high-precision numerical analysis and theoretical framework of QNMs in the language of non-Hermitian physics.
The resonance of QNMs not only allows us to decipher the long-standing mystery in black hole physics but also stands as a novel beacon in black hole spectroscopy.
Deviations from Kerr black holes, whether due to astrophysical effects or corrections to general relativity, will manifest through the resonances and be imprinted in the ringdown GWs.

Notably, similar enhancement effects around EPs in optics have already been employed to realize optical sensors with enhanced performance~\cite{Wiersig:20,Parto2021}. 
Additionally, it is well known that the Mikheyev-Smirnov-Wolfenstein resonance at the avoided crossing of neutrino eigenstates in matter plays a key role in solar neutrino spectroscopy~\cite{Smirnov:2003da,Wurm:2017cmm,Giganti:2017fhf}.
With high signal-to-noise ratio ringdown GWs from black holes with various spins expected in the near future, the resonant excitation of QNMs opens the door to rigorous examinations of black holes and the discovery of new physics in gravity.

\secl{Acknowledgments}
This work was supported by JSPS KAKENHI Grant No.~JP22K03639.

\secl{Data availability}
The data that support the findings of this article are openly available~\cite{motohashi_2024_12696858}.

\appendix
\renewcommand{\theequation}{S\arabic{equation}}
\setcounter{equation}{0}
\renewcommand{\thefigure}{S\arabic{figure}}
\setcounter{figure}{0}

\section*{Supplemental material}
\label{sec:supp}

In the main text we focus on the resonance at avoided crossing in the vicinity of EP of Kerr QNMs.
To demonstrate that this phenomenon is a universal feature of QNMs, it is helpful to study a simple toy model apart from Kerr QNMs.
Indeed, the simplest rectangular barrier potential played an important role in a pioneering work~\cite{Chandrasekhar:1975zza} in the early days of QNM research. 
Along the same line, here we consider a simple toy model 
\be \label{RWeq} \mk{\f{d^2}{dx^2} +\omega^2 -V(x)} \psi = 0 , \ee
with double rectangular barriers:
\be \label{dou-rec}
V(x) = 
\begin{cases}
V_0, & 0\leq x \leq d, \\
V_1, & b \leq x \leq b+d, \\
0, & {\rm otherwise},
\end{cases}
\ee
with $0\leq d\leq b$ and $V_0\geq 0$ and $V_1\geq 0$. 
We define the QNM frequencies by requiring the purely right-/left-going boundary condition, i.e., $\psi\to e^{\pm i\omega x}$, at $x>b+d$ and $x<0$, respectively. 
We can then calculate the excitation factor by \eqref{Bdef} or \eqref{Bana}. 
This toy model allows us to examine the intrinsic behavior of QNM frequencies and excitation factors in a simple and transparent manner.

Here, we introduce a hierarchy $V_1\ll V_0$ 
and investigate the behavior of QNM frequencies and excitation factors with respect to the variation of $b$.
For small values of $b$, the QNM spectrum is almost the same as the unperturbed one with $V_1=0$, but 
as we increase $b$, the QNM frequencies migrate on the complex frequency plane.  

The QNMs for a similar setup was investigated in \cite{Barausse:2014tra,Cheung:2021bol,Kyutoku:2022gbr} as the simplest variation of the ``flea on the elephant'' effect~\cite{Jona-Lasinio1981,Graffi_1984,SIMON1985123,Santarsiero_2019}, which has been explored recently in the context of the spectral instability~\cite{Nollert:1996rf,Barausse:2014tra,Daghigh:2020jyk,Jaramillo:2020tuu,Qian:2020cnz,Cheung:2021bol,Jaramillo:2021tmt,Berti:2022xfj,Kyutoku:2022gbr,Cardoso:2024mrw}.
It has been known that small deformation of potential at far distance causes a drastic change of QNM frequencies while changes in the time-domain ringdown signal show up only at late time.
However, the behavior of the excitation factors has not been clarified.

Let us classify the QNM frequencies for the double rectangular barriers~\eqref{dou-rec} into two categories:
The first sequence $\omega^{(1)}_n$ $(n=0,1,2,\cdots)$ denotes modes that are smoothly connected to the unperturbed modes with $V_1=0$.
The second sequence $\omega^{(2)}_k$ $(k=0,1,2,\cdots)$ denotes modes that show up when $V_1>0$ and correspond to the resonance frequencies for the effective cavity between the two barriers.
The labeling of overtone indices $n$ and $k$ is determined based on their decay times at $V_1 \to 0$.

For small $b$, the first sequence modes 
follow the logarithmic spiral around the unperturbed modes which is consistent with \cite{Leung:1997was}.
By using the perturbation theory formula~\eqref{dw-reg}, we obtain  
\be \omega^{(1)}_n \simeq \bar\omega_n -\f{i \bar B_n  V_1 \sin \bar\omega_n d }{\bar\omega_n} e^{i \bar\omega_n (2 b + d)} , \ee
where $\bar\omega_n$ and $\bar B_n$ are unperturbed QNM frequencies and excitation factors, respectively.
On the other hand, for the second sequence, we can derive a simple analytical expression under $V_1\ll V_0$ as
\be \label{QNM2} \omega^{(2)}_k \simeq \f{1}{2b} \kk{ (2k+1) \pi - i \ln \f{V_0}{V_1} } . \ee
Thus, as we increase $b$, the second sequence modes accumulate to the origin of the complex frequency plane, during which the second sequence modes (or migrating modes more generally, including modes that were originally belonging to the first sequence) approach to the first sequence modes, which then begin to migrate together with the migrating modes.

\begin{figure*}[t]
\centering
\includegraphics[width=.3\textwidth]{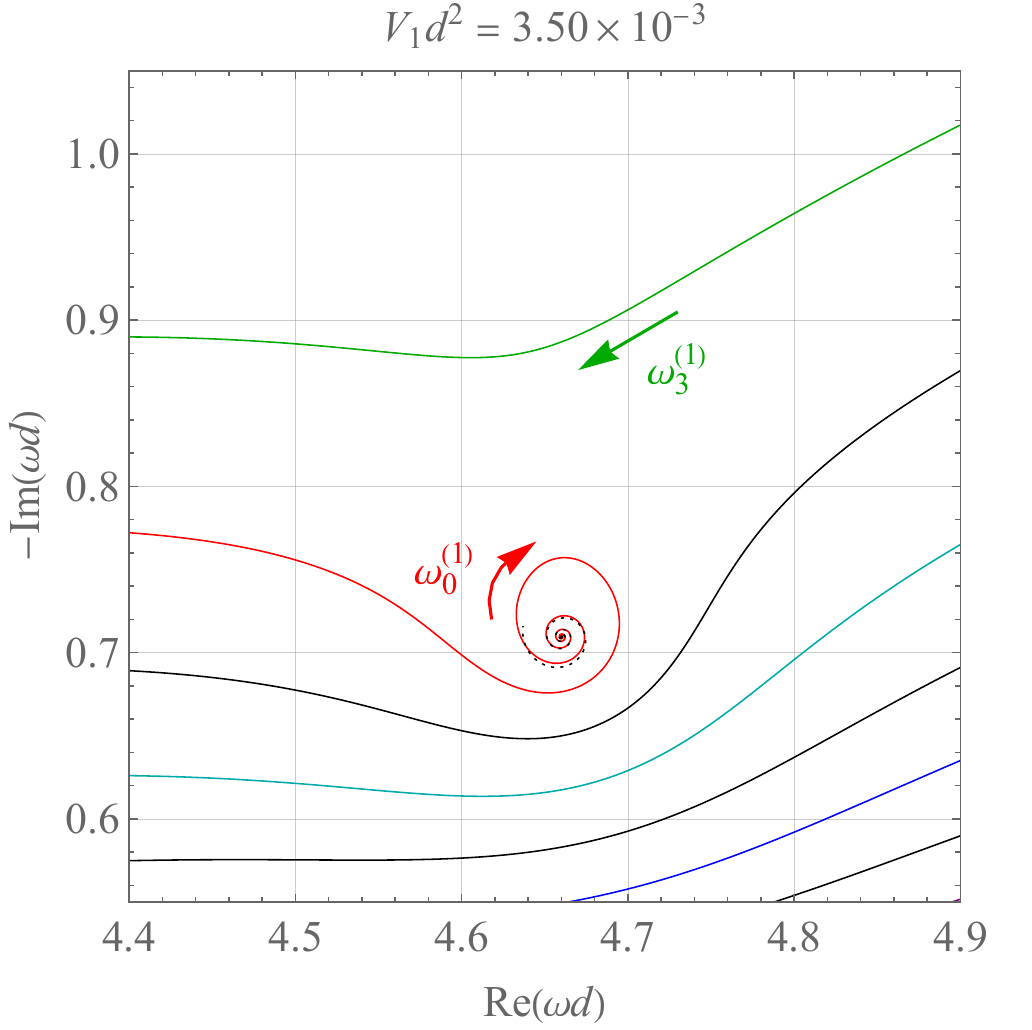}
\includegraphics[width=.317\textwidth]{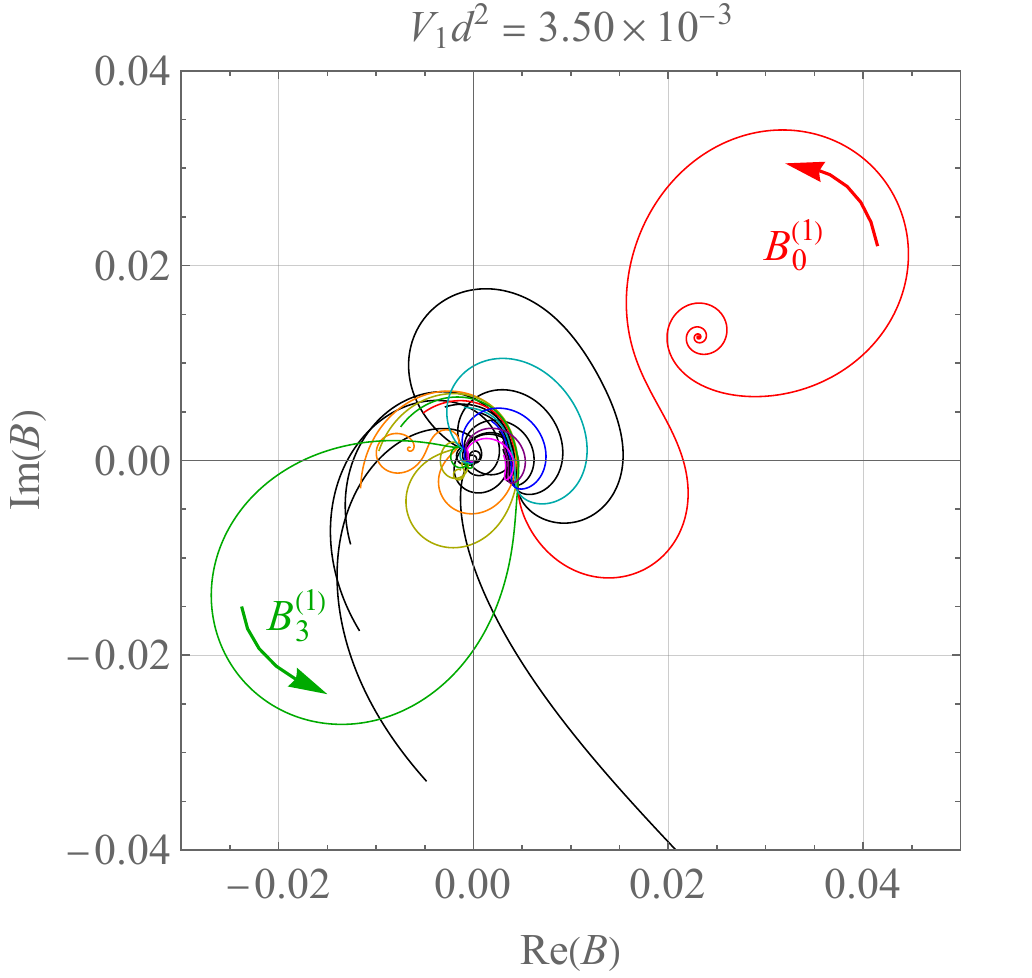}
\includegraphics[width=.35\textwidth]{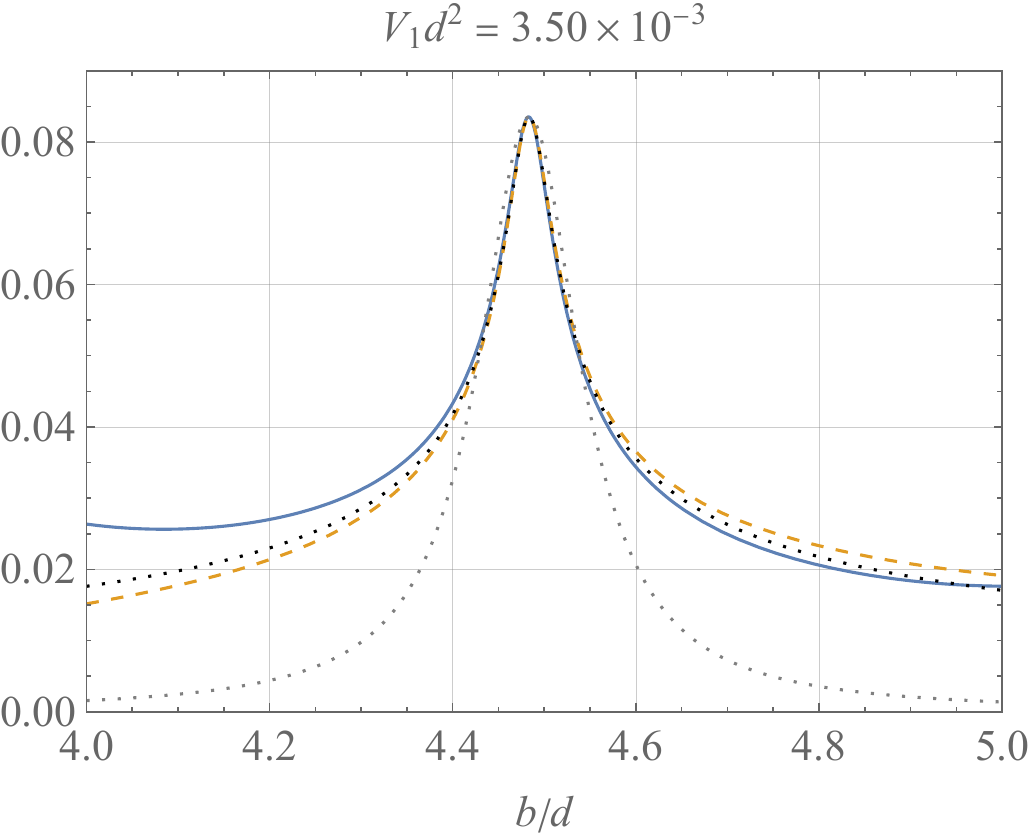}
\includegraphics[width=.3\textwidth]{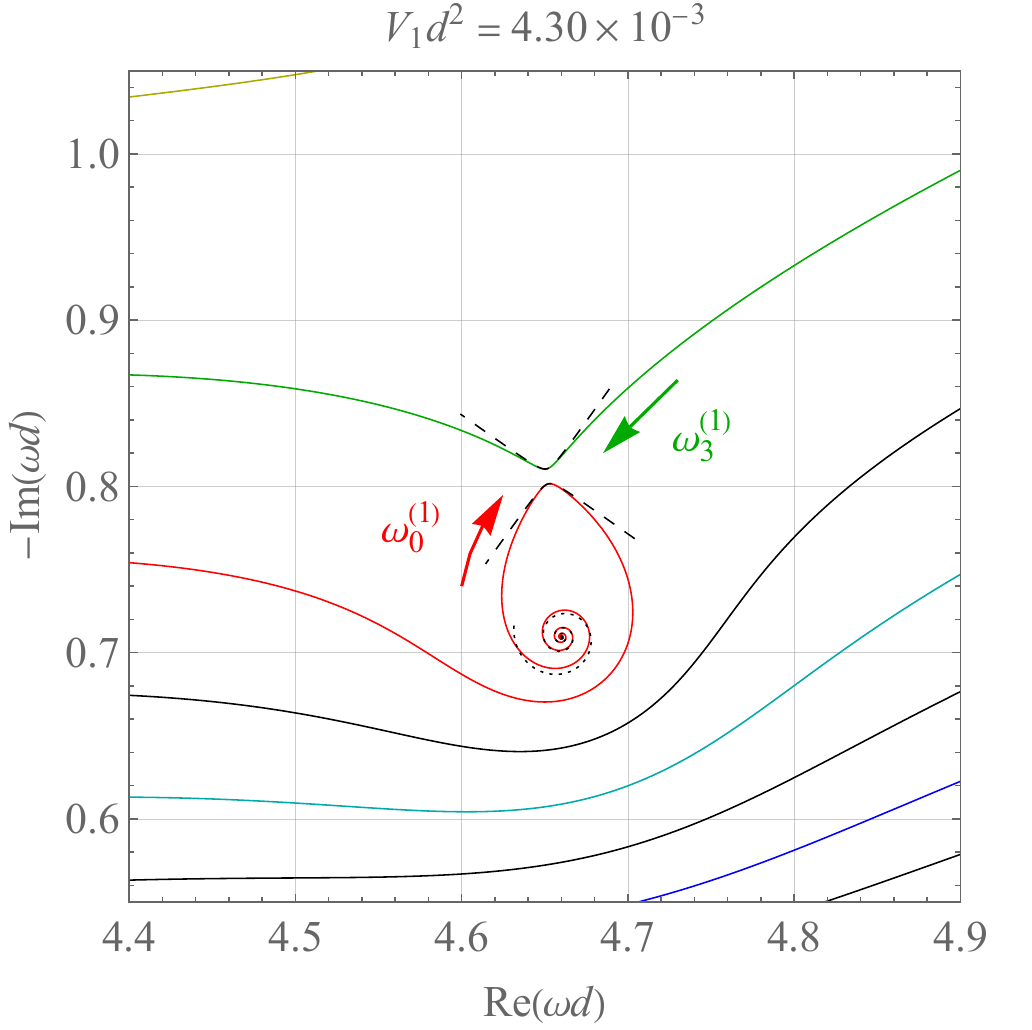}
\includegraphics[width=.312\textwidth]{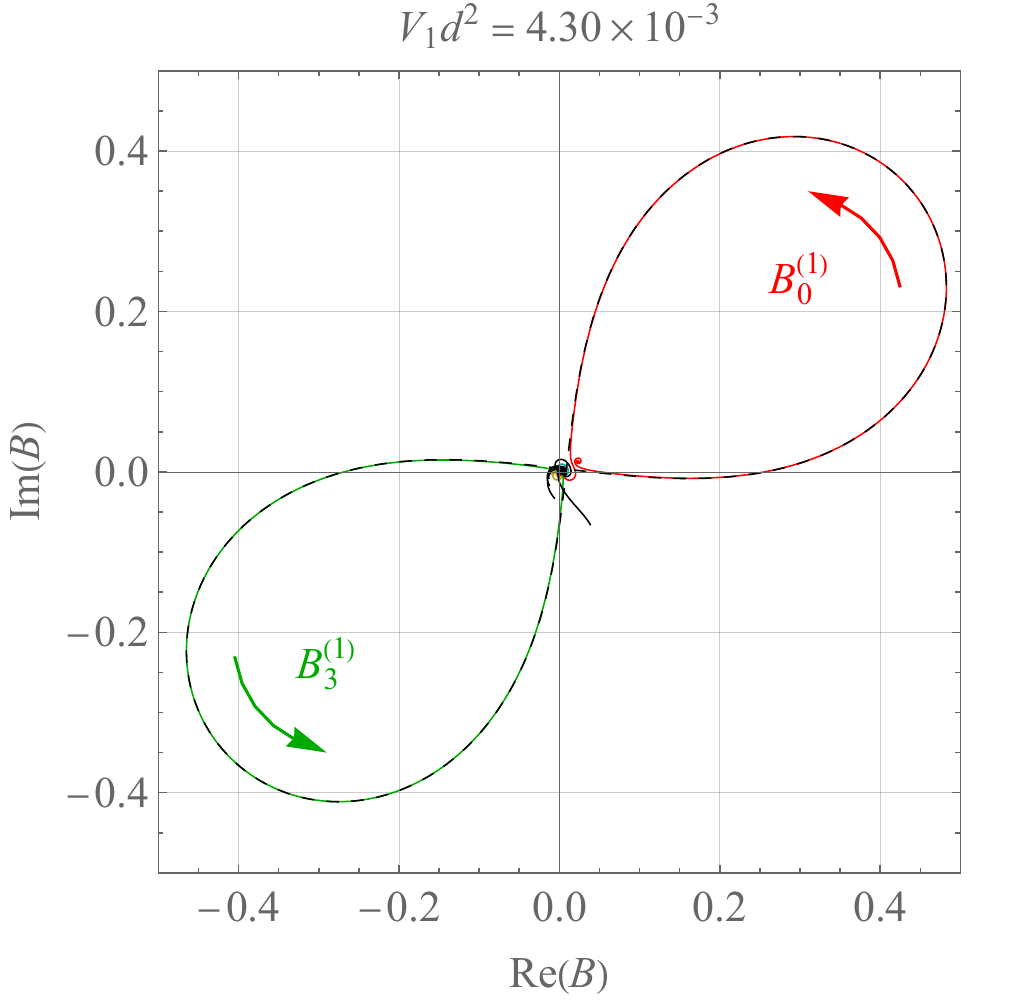}
\includegraphics[width=.35\textwidth]{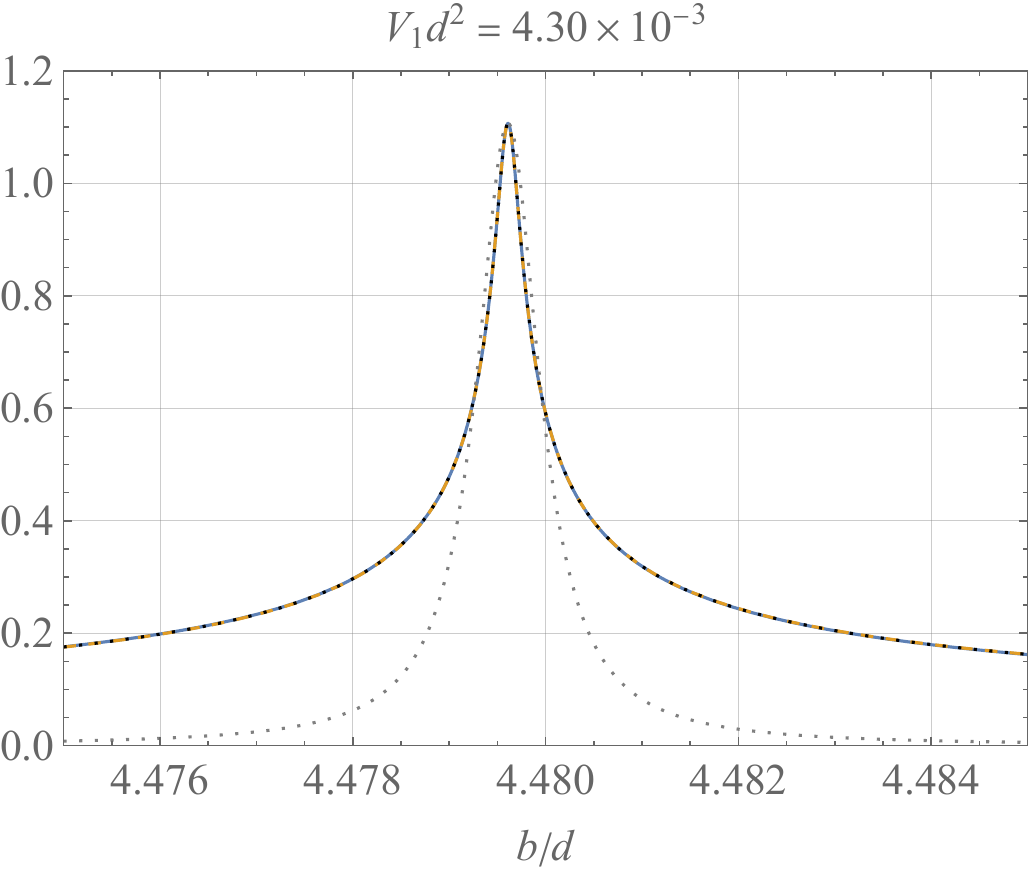}
\includegraphics[width=.3\textwidth]{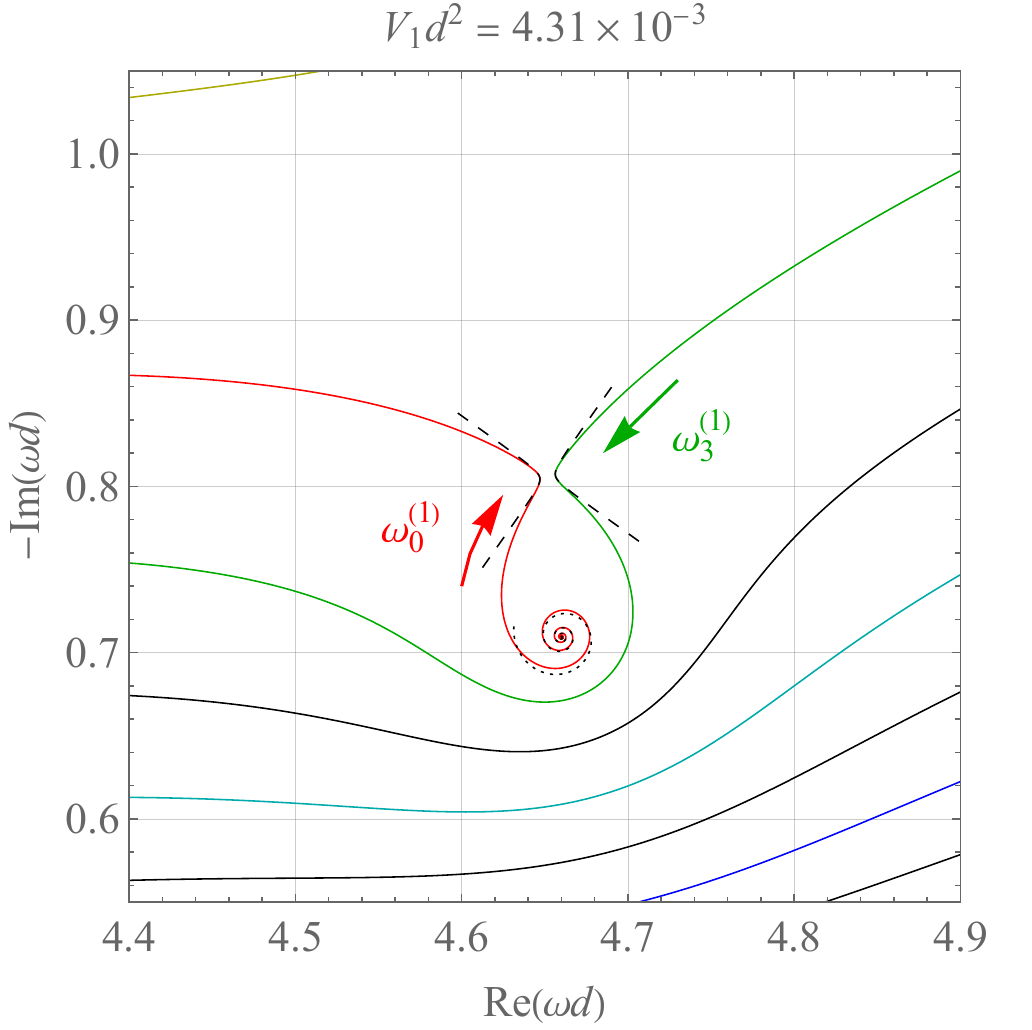}
\includegraphics[width=.312\textwidth]{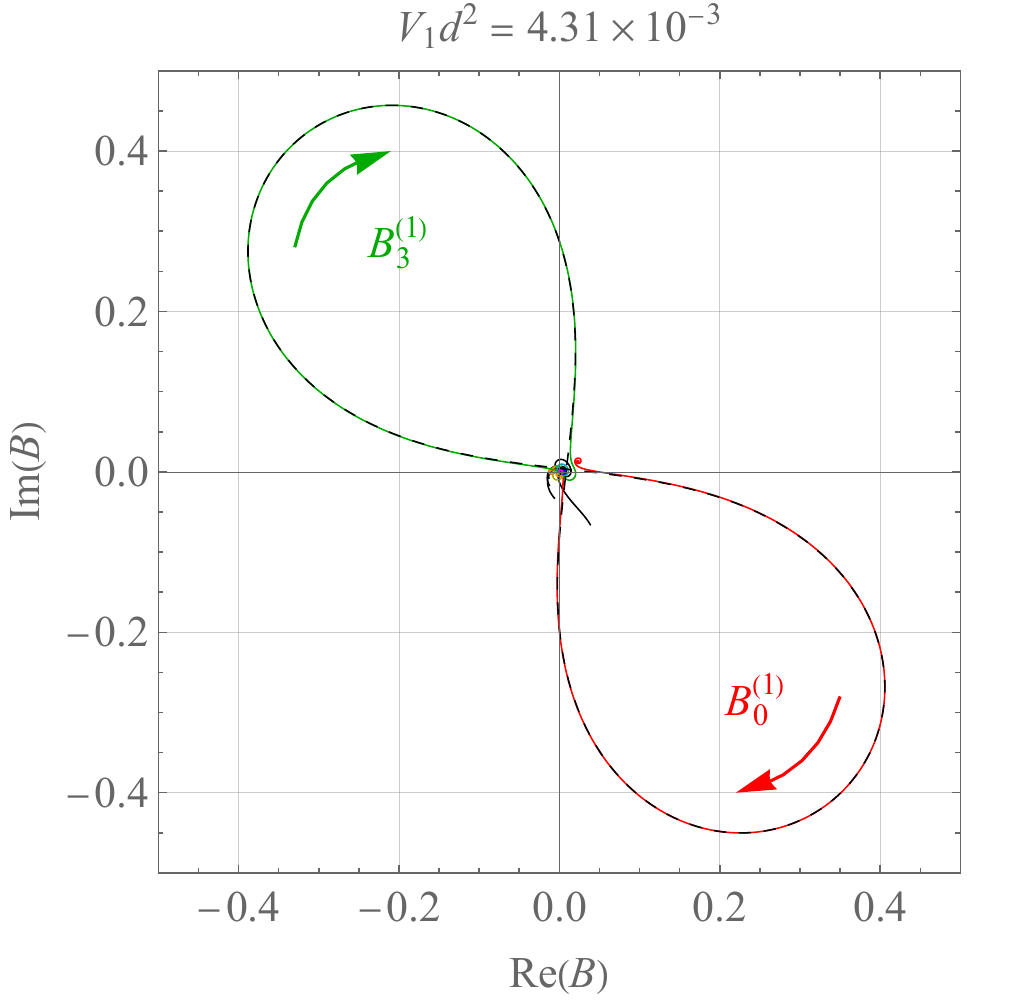}
\includegraphics[width=.35\textwidth]{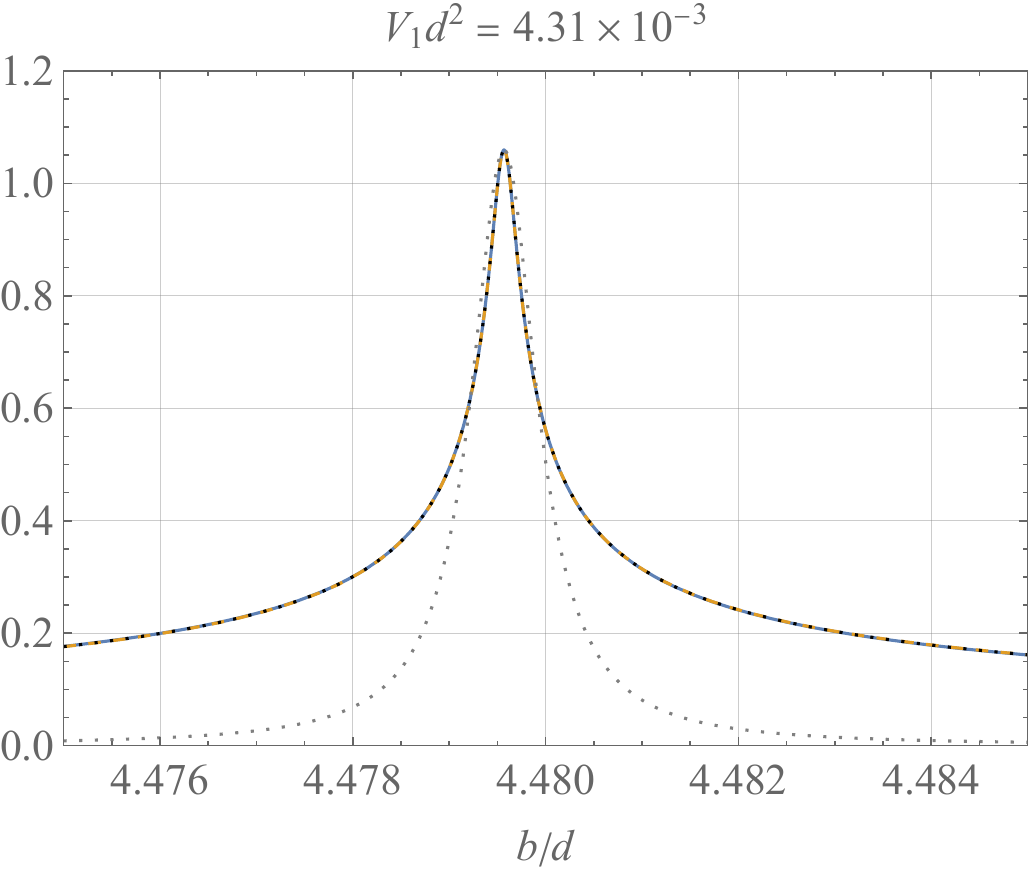}
\includegraphics[width=.3\textwidth]{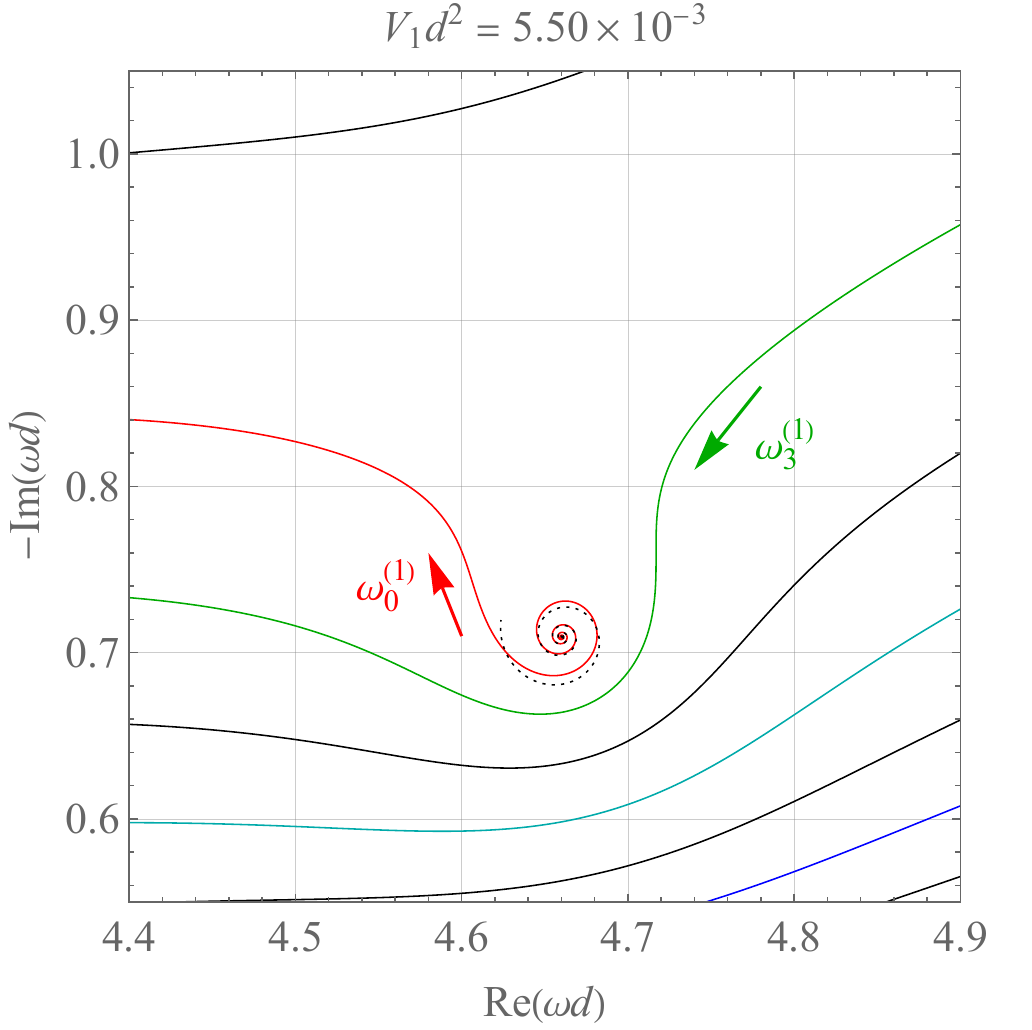}
\includegraphics[width=.317\textwidth]{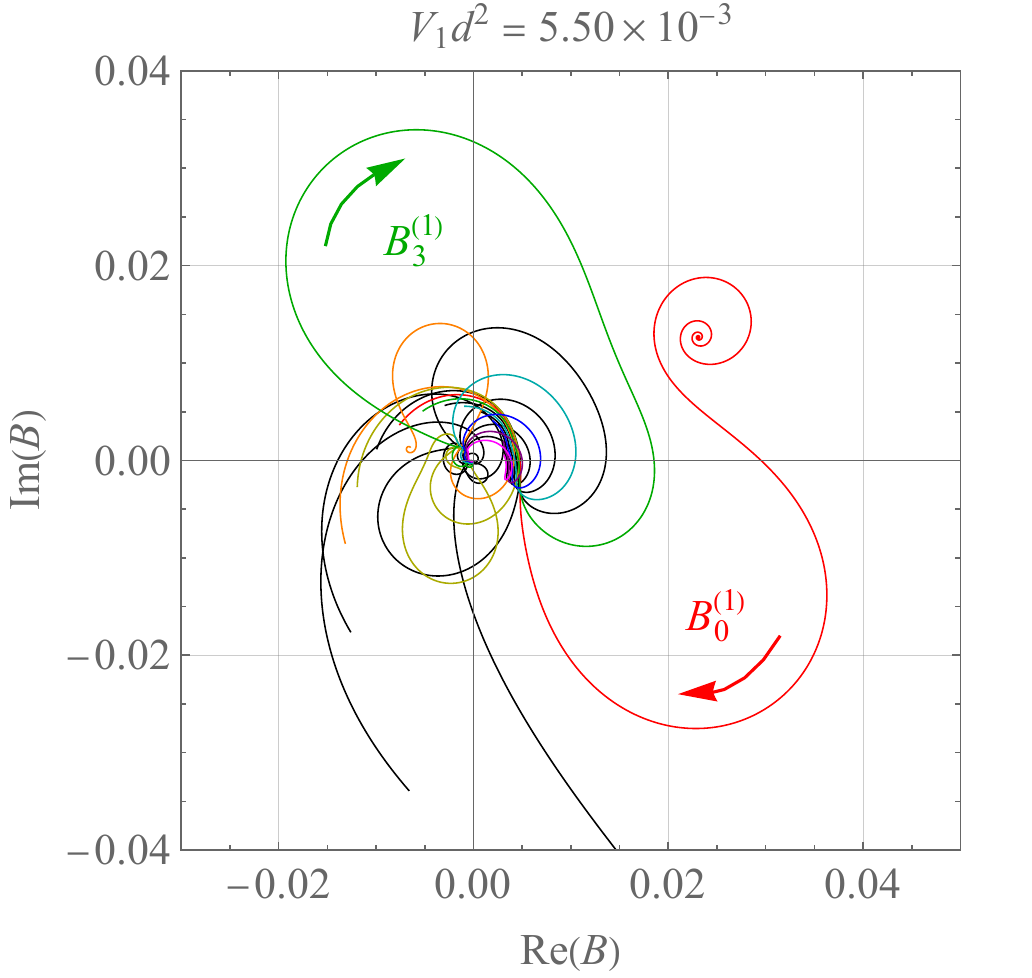}
\includegraphics[width=.35\textwidth]{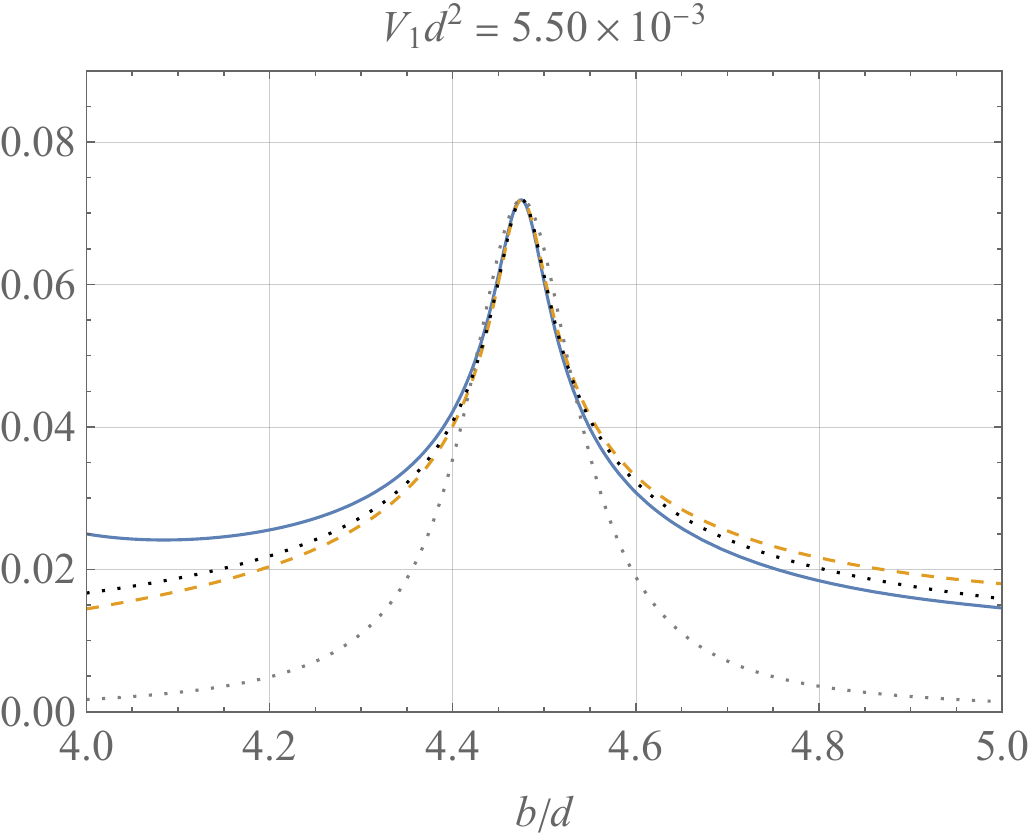}
\caption{{\it Left and middle columns:} 
QNM frequencies (left column) and excitation factors (middle column) for the double rectangular barriers~\eqref{dou-rec}.
The fundamental mode $\omega^{(1)}_0$ (red) and third overtone $\omega^{(1)}_3$ (green) of the first sequence are highlighted, while overtones up to the seventh in the first (colored) and second (black) sequences are depicted. 
We set parameters $V_0d^2=16$ and $V_1d^2=(3.50,4.30,4.31,5.50)\times 10^{-3}$ (from first to fourth row) and depict trajectories with respect to the variation of $b/d$ in the range $1\leq b/d\leq 15$ with arrows indicating the direction of increasing $b/d$.  
The logarithmic spiral (black dotted), hyperbola (black dashed), and lemniscate of Bernoulli (black dashed) approximate the trajectories well. 
{\it Right column:}
$|B^{(1)}_0 - B^{(1)}_3|$ (blue solid) and $1/|\omega^{(1)}_0-\omega^{(1)}_3|$ (orange dashed), the latter of which is multiplied by a numerical constant to match the maxima.  
The peak can be well fitted by a quarter power of Lorentzian (black dotted) rather than the Lorentzian itself (gray dotted).}
\label{fig:DR}
\end{figure*}

We find that when two QNM frequencies approach each other during the migration, they exhibit the same behavior observed for the Kerr QNMs:
The two QNM frequencies repel each other, during which the excitation factors are point-symmetrically amplified.
In the left and middle columns of Fig.~\ref{fig:DR}, we depict the trajectories of QNM frequencies and excitation factors with respect to the variation of $b$.
In addition to $b$, we control the second parameter $V_1$ to see how the avoided crossing and resonance change.
The cases with $V_1d^2=4.30\times 10^{-3}$ and $4.31\times 10^{-3}$ demonstrate sharp resonances, where the trajectories follow the hyperbola and lemniscate. 
The two cases exhibit a 90-degree switch in the complex plane,
which is predicted at the sign flip of the RHS of $X^2-Y^2 = q^2\sin 2(\alpha-\beta)$ in the main text.
The cases with $V_1d^2=3.50\times 10^{-3}$ and $5.50\times 10^{-3}$ demonstrate mild resonances, where the trajectories of the QNM frequencies and excitation factors can be regarded as distorted versions of the hyperbola and lemniscate, respectively.
These sharp and mild resonances reproduce what we observed in the Kerr QNM frequencies and excitation factors in the main text.

We also examine the characteristic resonant peak appearing in the difference of the excitation factors and the inverse of the difference of the QNM frequencies.
As shown in the right column of Fig.~\ref{fig:DR}, the peaks are well fitted by the quarter-power Lorentzian, same as the Kerr case.
For the sharp resonances, the relative difference between any pair of $\Delta B$, $1/\Delta \omega$ (normalized by multiplying a numerical constant), and the quarter-power Lorentzian remains $\mO(1)$\% in the plotted region.
The mild resonances are still better fitted by the quarter-power Lorentzian over the Lorentzian itself.

Note also that the resonance phenomenon occurs between the fundamental mode and third overtone of the first sequence.
This reveals that, while the resonances for Kerr QNMs studied in the main text primarily involve higher overtones, the resonances involving the fundamental mode and/or lower overtones are in general not prohibited.
Indeed, we analyze the double rectangular barriers across various parameter sets and confirm the resonances for various pairs of modes approaching each other.
This supports that the resonance is a universal feature of QNMs. 

\clearpage

\bibliographystyle{apsrev4-1mod}
\bibliography{ref}

\end{document}